\documentclass[dvips,twoside,fleqn]{article}
\usepackage{times}
\usepackage{epsfig}
\usepackage{amstext}
\usepackage{amssymb}
\usepackage{espcrc2}

\title{Exploring the Unstable Modes Dynamics by the Lattice
        Schr\"odinger Functional}

\author{
Paolo Cea\address{Dipartimento di Fisica - Universit\`a di Bari,
               via Amendola 173,
               70126 Bari, Italy}$^{\text{,b}}$
and Leonardo Cosmai\address{INFN - Sezione di Bari,
               via Amendola 173, 70126 Bari, Italy}
 }

\begin{document}


\begin{abstract}
\noindent
We analyze the problem of the Nielsen-Olesen unstable modes in
the $SU(2)$ lattice gauge theory by means of a recently introduced
gauge-invariant effective action. We perform numerical simulations
in the case of a constant Abelian chromomagnetic field. We find
that for lattice sizes above a certain critical length the density
of effective action shows a behaviour compatible with the presence
of the unstable modes.
\end{abstract}

\maketitle

\section{The Lattice Effective Action}

The Euclidean Schr\"odinger functional in Yang-Mills theories without
matter fields is:
\begin{equation}
\label{Eq1}
{\mathcal{Z}} \left[ A^{(f)}, A^{(i)} \right] =
 \langle  A^{(f)}  |  \exp(-HT) {\mathcal{P}} |
A^{(i)}
\rangle
\end{equation}
where ${\mathcal{P}}$ projects onto the physical
states and $H$ is the Hamiltonian of the gauge system.

On the lattice the Schr\"odinger functional reads:
\begin{equation}
\label{Eq2}
{\mathcal{Z}} \left[ U^{(f)}, U^{(i)} \right] =
\int {\mathcal{D}}U \exp(-S)
\,,
\end{equation}
where $S$ is the Wilson action and we impose the boundary conditions:
\begin{equation}
\label{Eq3}
U(x)|_{x_4=0} = U^{(i)} \, ,  \quad  U(x)|_{x_4=T} = U^{(f)} \,.
\end{equation}
Note that the Schr\"odinger functional is 
{\em invariant} under arbitrary lattice gauge
transformations of the boundary links.
If we consider :
\begin{equation}
\label{Eq4}
U(x)|_{x_4=0} = U(x)|_{x_4=T} = U^{\text{ext}}(0,\vec{x}) \,.
\end{equation}
where the lattice links are related to the continuum gauge fields
 by the well-known relation (${\text{P}}$ is the path-ordering operator):
\begin{equation}
\label{Eq5}
U_\mu^{\text{ext}}(x) = {\text{P}} \exp \left\{ + iag  \int_0^1 dt \,
A_\mu^{\text{ext}}(x+ at {\hat{\mu}}) \right\} \, ,
\end{equation}
then we define~\cite{Cea1} the lattice effective action for 
the background field
$A_\mu^{\text{ext}}(\vec{x})$ by means of the  lattice
Schr\"odinger functional Eq.~(\ref{Eq2}) as:
\begin{equation}
\label{Eq7}
\Gamma\left[ \vec{A}^{\text{ext}} \right] = -\frac{1}{T}
\ln \left\{ \frac{{\mathcal{Z}}[U^{\text{ext}}]}{{\mathcal{Z}}(0)} \right\} \,.
\end{equation}
In Eq.~(\ref{Eq7}) T is the extension in Euclidean time,
${\mathcal{Z}}[U^{\text{ext}}]=
{\mathcal{Z}}[U^{\text{ext}},U^{\text{ext}}]$ and  ${\mathcal{Z}}(0)$ is
is the lattice Schr\"odinger functional
without external background field  ($U_\mu^{\text{ext}} =1$).

Our effective action is by definition gauge invariant
and can be used for a non-perturbative
investigation of the properties of the quantum vacuum.
 In the case of background fields which  give rise to a  constant field
strength
 $\Gamma\left[ \vec{A}^{\text{ext}} \right]$ is
proportional to the spatial volume  $V$ . Thus we are interested in the
density of the effective action:
\begin{equation}
\label{Eq8}
\varepsilon\left[ \vec{A}^{\text{ext}} \right] =
-\frac{1}{\Omega} \ln \left[
\frac{{\mathcal{Z}}[U^{\text{ext}}]}{{\mathcal{Z}}(0)} \right] 
\,, \quad \Omega=V \cdot T \,.
\end{equation}
Our proposal for the gauge invariant effective action on the lattice
has been successfully checked in the U(1) lattice gauge 
theory~\cite{Cea1,Cea2,Cea3}.

Let us consider the $SU(2)$ lattice gauge theory in a constant 
Abelian background magnetic field:
\begin{equation}
\label{Eq9}
\vec{A}_a^{\text{ext}}(\vec{x}) =\vec{A}^{\text{ext}}(\vec{x})\delta_{a,3} \,
 , \, 
A_k^{\text{ext}}(\vec{x}) = \delta_{k,2} x_1 H \,.
\end{equation}
On the lattice:
\begin{eqnarray}
\label{Eq10}
&&  U^{\text{ext}}_2(x) = \cos(\frac{agHx_1}{2})  + i \sigma^3 
 \sin(\frac{agHx_1}{2})
\nonumber \\
&&  U^{\text{ext}}_1(x) =  U^{\text{ext}}_3(x) = U^{\text{ext}}_4(x) = 1 
\,.
\end{eqnarray}
The periodic boundary conditions imply the  quantization of the
external magnetic field
\begin{equation}
\label{Eq11}
\frac{a^2 g H}{2} = \frac{2 \pi}{L_1} n^{\text{ext}}
\end{equation}
where $n^{\text{ext}}$ is an integer,  $L_1$ the lattice extension in the 
$x_1$ direction (in lattice units).

We perform numerical simulations by using the standard
Wilson action.  The links belonging to the time slice $x_4=0$ are
frozen to the configuration Eq.~(\ref{Eq10}). We also impose that the
constraints Eq.~(\ref{Eq10}) apply to links at the spatial boundaries
(in the continuum this condition amounts to the usual requirement that
the fluctuations over the background field vanish at the infinity).
In order to evaluate the density of the effective action
Eq.~(\ref{Eq7}) we are faced with the 
problem of computing a  partition function. 
To avoid this problem we consider the derivative of
$\varepsilon[\vec{A}^{\text{ext}}]$
with respect to $\beta$:
\begin{eqnarray}
\label{Eq12}
 \varepsilon^{\prime} \left[ \vec{A}^{\text{ext}} \right] =
&& \left \langle \frac{1}{\Omega} 
 \sum_{x,\mu>\nu} \frac{1}{2} \text{Tr}
 U_{\mu\nu}(x) \right \rangle_0 \, -
\nonumber \\
&&  
 \left \langle \frac{1}{\Omega} \sum_{x,\mu>\nu} \frac{1}{2} \text{Tr}
U_{\mu\nu}(x) \right \rangle_{A^{\text{ext}}} \,.
\end{eqnarray}
\section{The Unstable Modes}

Let us briefly discuss the origin of the Nielsen-Olesen unstable 
modes~\cite{Nielsen78}.
 To evaluate the continuum effective action one writes:
\begin{equation}
\label{Eq13}
A(x) =  A^{\text{ext}}(x) + \eta(x)
\end{equation}
where $ \eta(x)$ is the quantum fluctuation on the background field.
In the one-loop approximation one retains in the action only the terms 
 quadratic in 
 $ \eta(x)$ . After performing the Gaussian functional
 integration in the background gauge one obtains~\cite{Savvidy77}: 
\begin{equation}
\label{Eq:14}
 \varepsilon(H)  = \frac{1}{2} H^2 + \frac{11}{48 \pi^2} g^2H^2
 \ln(\frac{gH}{\Lambda^2}) + {\cal O}(g^2H^2)
\end{equation}
 However, it was pointed out~\cite{Nielsen78} that the quadratic action
 is affected by the presence of negative eigenvalues which give rise
 to the unstable modes.
 It turns out that the stabilization of the unstable modes induces
 a negative classical-like term~\cite{Consoli85} which cancels
 the classical energy~\cite{Cea4}:
\begin{equation}
\label{Eq15}
 \varepsilon(H)  =  \frac{1}{48 \pi^2} g^2H^2
 \ln(\frac{\Lambda^2}{gH}) + {\cal O}(g^2H^2) 
\end{equation}
in the thermodynamic limit $V \rightarrow \infty$.

On the lattice we must evaluate the effective action in the weak coupling
region. In the one-loop approximation it turns out that even the lattice
quadratic action displays the unstable modes. Discarding some irrelevant
operators we find an approximate formula for the unstable modes
eingenvalues:
\begin{equation}
\label{Eq16}
\lambda_u \, = \, (1 - \cos p_4) + (1 - \cos p_3) - \sin(\frac{gH}{2}) \;,
\end{equation}
where $ p_\mu =  \frac{2 \pi}{L_\mu} n_\mu$ .
In our simulations we fix $L_3=L_4=32$ and vary $L_1=L_2=L$ with 
$n^{\text{ext}}=1$. Inserting these values into Eq.~(\ref{Eq16}) we
find $\lambda_u \leq 0$ when $ L \geq L_{\text{crit}} \simeq 10$.
So that we can switch on and off the unstable modes on the 
lattice~\cite{Levi95}.
\begin{figure}[t]
\label{Fig1}
\begin{center}
\epsfig{file=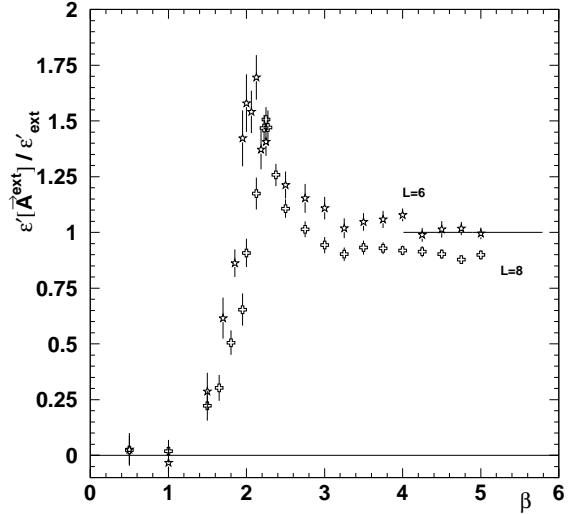,width=7.5truecm}
\caption{ $\frac{\varepsilon^{\prime}[ \vec{A}^{\text{ext}} ]}
{ \varepsilon^{\prime}_{\text{ext}} }$ versus $\beta$
for $L=6$ (stars) and $L=8$ (crosses).}
\end{center}
\end{figure}
\begin{figure}[t]
\begin{center}
\label{Fig2}
\epsfig{file=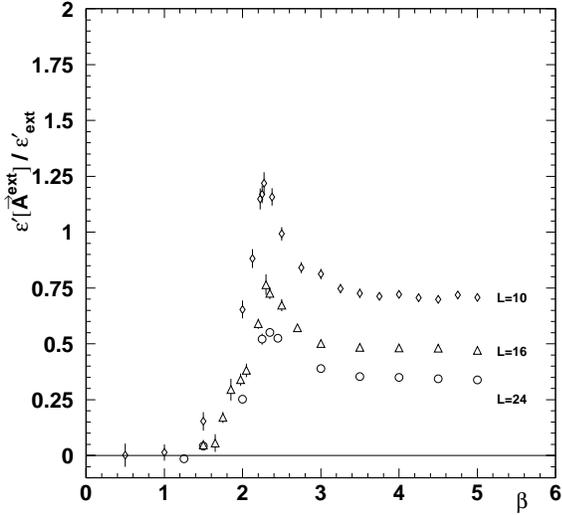,width=7.5truecm}
\caption{ $\frac{\varepsilon^{\prime}[ \vec{A}^{\text{ext}} ]}
{ \varepsilon^{\prime}_{\text{ext}} }$ versus $\beta$
for $L=10$ (diamonds), $L=16$ (triangles) and $L=24$ (circles).}
\end{center}
\end{figure}
\begin{figure}[t]
\label{Fig3}
\begin{center}
\epsfig{file=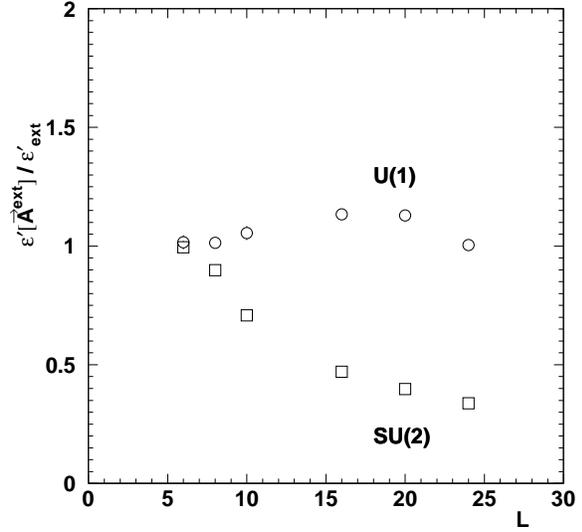,width=7.5truecm}
\caption{The ratio $\frac{\varepsilon^{\prime}[ \vec{A}^{\text{ext}} ]}
{ \varepsilon^{\prime}_{\text{ext}} }$ in the weak coupling region
 versus $L$ for $U(1)$ (circles) and $SU(2)$ (squares).}
\end{center}
\end{figure}

In Figs.~1 and~2 we show the derivative of the
density of the effective action ( in units of 
$\varepsilon^{\prime}_{\text{ext}}=1- \cos ( \frac{2 \pi}{L_1}
 n^{\text{ext}} ) \, $ ) for different lattice sizes.
We see that in the weak coupling region 
 $\varepsilon^{\prime}[ \vec{A}^{\text{ext}} ]$ tends to the derivative 
of the external action if $L \leq 8$. On the other hand, if
$L > 8$  the derivative of the density of the effective action 
 decreases monotonously by increasing the lattice volume.
 This peculiar behaviour is a truly dynamic effect due to
 the unstable modes. Indeed  Fig.~3,  where 
 we contrast the $U(1)$ and $SU(2)$  case in the weak coupling
 region ( $\beta = 3 $ for $U(1)$ and $\beta = 5 $ for $SU(2)$ ),
 shows that in the $U(1)$ theory 
$\frac{\varepsilon^{\prime}[ \vec{A}^{\text{ext}} ]}
{ \varepsilon^{\prime}_{\text{ext}} } \simeq 1$.
 Moreover, according to Eq.~(\ref{Eq15}) we find that in the
 weak coupling region ($\beta = 5$):
\begin{equation}
\label{Eq17}
 \frac{\varepsilon^{\prime}[ \vec{A}^{\text{ext}} ]}{\varepsilon^{\prime}_{
 \text{ext}}} \; = \;
 \frac{a}{L_{\text{eff}}^\alpha} \; \; ; \; \alpha = 1.47(7) \, ,
\end{equation}
where $L_{\text{eff}}$ is the effective linear size of the 
lattice~\cite{Cea2}.
Remarkably, it turns out that also the peak value of 
 $\varepsilon^{\prime}[ \vec{A}^{\text{ext}}]$ tends towards
 zero with the same law as implied by Eq.~(\ref{Eq17}):
\begin{equation}
\label{Eq18}
 \frac{\varepsilon^{\prime}[ \vec{A}^{\text{ext}} ]}{\varepsilon^{\prime}_{
 \text{ext}}} \,|_{\text{peak}} \, = \,
 \frac{a^{\prime}}{L_{\text{eff}}^{\alpha^{\prime}}} \; \; ; \; 
 \alpha^{\prime} = 1.5(1) \, .
\end{equation}
Equations (~\ref{Eq17}) and ( ~\ref{Eq18}) tell us that 
$\varepsilon^{\prime}[ \vec{A}^{\text{ext}}]$
tends uniformly towards zero in the thermodynamic limit.
\section{Conclusions}

Our numerical results are suggesting that 
$\varepsilon^{\prime} \left[ \vec{A}^{\text{ext}} \right]
\rightarrow 0$ when $L_{\text{eff}} \rightarrow \infty$ in the whole
$\beta$-region. Thus in the continuum limit
$L_{\text{eff}} \rightarrow \infty$, $\beta \rightarrow \infty$
the confining vacuum screens completely the external chromomagnetic Abelian
field. In other words, the continuum vacuum behaves as an Abelian
 magnetic condensate medium in accordance with the dual superconductivity 
scenario. Moreover the magnetic condensate dynamics seems to be closely
related to the presence of the Nielsen-Olesen unstable modes.
Thus, our approach to a gauge-invariant effective action on the
lattice opens the door towards the understanding of the dynamics of
color confinement.

\end{document}